\documentclass[]{spie}  

\usepackage{amsmath,amsfonts,amssymb}
\usepackage{graphicx}
\usepackage[colorlinks=true, allcolors=blue]{hyperref}
\usepackage{aas_macros}

\title{High-contrast imager for complex aperture telescopes (HiCAT): 11. System-level demonstration of the Apodized Pupil Lyot Coronagraph with a segmented aperture in air}

\author[a]{R\'emi Soummer}
 \author[a]{Raphaël Pourcelot}
 \author[a]{Emiel H. Por}
 \author[a]{Sarah Steiger}
 \author[d]{Iva Laginja}
 \author[b]{Benjamin Buralli}
 \author[k]{Susan Redmond}
 \author[a]{Laurent Pueyo}
 \author[a]{Marshall D. Perrin}
 \author[g]{Marc Ferrari}
 \author[l]{Jules Fowler}
 \author[h]{John Hagopian}
 \author[b]{Mamadou N'Diaye}
 \author[a]{Meiji Nguyen}
 \author[a]{Bryony Nickson}
 \author[a,e]{Peter Petrone}
 \author[a]{Ananya Sahoo}
 \author[a]{Anand Sivaramakrishnan}
 \author[e]{Scott D. Will}

 \affil[a]{Space Telescope Science Institute, 3700 San Martin Drive, Baltimore, USA}
 \affil[b]{Université Côte d’Azur, Observatoire de la Côte d’Azur, CNRS, Laboratoire Lagrange, France }
 \affil[d]{LESIA, Observatoire de Paris, Université PSL, Sorbonne Université, Université Paris Cité, CNRS, 5 place Jules Janssen, 92195 Meudon, France}
 \affil[e]{NASA Goddard Space Flight Center, Greenbelt, MD 20771, USA}
 \affil[g]{Aix Marseille Université, CNRS, CNES, LAM (Laboratoire d’Astrophysique de Marseille) UMR 7326, 13388 Marseille, France}
 \affil[h]{Advanced Nanophotonics Inc., 4437 Windsor Farm Rd, Harwood, MD USA}
 \affil[k]{Caltech and Jet Propulsion Laboratory, Pasadena, CA, USA}
 \affil[l]{University of California Santa Cruz, Santa Cruz, CA, USA}

\authorinfo{Further author information, send correspondence to R.S.: E-mail: soummer@stsci.edu}

\begin{document} 
\maketitle

\begin{abstract}

We present the final results of the Apodized Pupil Lyot Coronagraph (APLC) on the High-contrast imager for Complex Aperture Telescopes (HiCAT) testbed,  under NASA's Strategic Astrophysics Technology program.  The HiCAT testbed was developed over the past decade to enable a system-level demonstration of coronagraphy for exoplanet direct imaging with the future Habitable Wolds Observatory.  HiCAT includes an active, segmented telescope simulator, a coronagraph, and metrology systems (Low-order and Mid-Order Zernike Wavefront Sensors, and Phase Retrieval camera). These results correspond to an off-axis (un-obscured) configuration, as was envisioned in the 2020 Decadal Survey Recommendations.  Narrowband and broadband dark holes are generated using two continuous deformable mirrors (DM) to control high order wavefront aberrations, and low-order drifts can be further stabilized using the LOWFS loop.  The APLC apodizers, manufactured using carbon nanotubes, were optimized for broadband performance and include the calibrated geometric aperture.  

The objectives of this SAT program were organized in three milestones to reach a system-like level demonstration of segmented-aperture coronagraphy, from static component demonstration to system-level demonstration under both natural and artificial disturbances.   HiCAT is, to this date, the only testbed facility able to demonstrate high-contrast coronagraphy with a truly segmented aperture, as is required for the Habitable World Observatory, albeit limited to ambient conditions, corresponding to NASA's Technology Readiness Level (TRL) 4. Results presented here include $6\times 10^{-8}$ (90\% CI) contrast in 9\% bandpass in a 360 deg dark hole with inner and outer working angles of $4.4 \lambda/D_{pupil}$ and $11 \lambda/D_{pupil}$ .   Narrowband contrast (3\% bandpass) reaches $2.4\times 10^{-8}$ (90\% confidence interval).  
We first explore the open-loop stability of the entire system quantify the baseline testbed performance.  Then we present dark hole stabilization using both high-order and low-order loops under both low-order and segment level drifts in narrow and broadband.  We compare experimental data with that obtained by the end-to-end HiCAT simulator.  We establish that current performance limitations are due to a combination of ambient conditions, detector and deformable mirrors noises (including quantization), and model mismatch.  

\end{abstract}


\section{INTRODUCTION}

NASA’s Habitable Worlds Observatory (HWO) is being developed to seek out signs of life on $\sim 25$ Earth-like planets orbiting Sun-like stars. This historic search for life outside our Solar System challenges us to build a coupled telescope and coronagraph system capable of suppressing starlight at $\sim 10^{-10}$ contrast so that a  $\sim 6$ m HWO can search $\sim 100$ star systems for Earth-like planets. 

NASA’s Roman Space Telescope will prove some of the key technologies needed for HWO\cite{Bailey2023SPIE}, but will not be able to detect and characterize Earth-sized worlds. The LUVOIR and HabEx studies \cite{theluvoirteam2019luvoirmissionconceptstudy,gaudi2020habitableexoplanetobservatoryhabex} detailed in depth the technology needed to  achieve $10^{-10}$ contrast to detect planets and take deep spectra that might signal life. Recognizing these challenges, the Astro2020 Decadal Survey \cite{NAP26141} recommended a Great Observatories Maturation Program (GOMAP) to mature the technologies needed to make HWO feasible as a segmented telescope, and a HWO Program Office is being created to lead  mission development. 

In light of the exquisite optical performance and stability of JWST, NASA has indicated it will adopt this high-heritage segmented format for HWO. Segmented systems require the telescope and coronagraph to be seamlessly coupled.   The High-contrast Imager for Complex Aperture Telescopes (HiCAT), started in 2013, has been developed to advance the Technology Readiness Level (TRL) of high-contrast imaging with segmented apertures in space \cite{ndiaye2013SPIE,ndiaye2014SPIE_hicat2,2015SPIE.9605E..0IN,2016SPIE.9904E..3CL,2017SPIE10562E..2ZL,2018SPIE10698E..53M,2018SPIE10698E..1OS}. HiCAT was designed to achieve a first system-like demonstration with a  segmented input telescope simulator, an optimized coronagraph, and control software for continuous contrast maintenance.  The testbed also includes metrology capabilities (Low and Mid-order wavefront sensor, phase retrieval camera) and can implement multiple coronagraphic modes: Classical Lyot Coronagraph (CLC) \cite{2022SPIE12180E..26S} Apodized Pupil Lyot Coronagraph (APLC)\cite{2005ApJ...618L.161S, 2015ApJ...799..225N, ndiaye2016, 2016JATIS...2a1012Z} and Phase-Apodized-Pupil Lyot Coronagraph (PAPLC)\cite{Por2020PAPLC}. 

 In this paper, we report the latest results with the APLC coronagraph in both narrowband and broadband with updated designs combining an APLC with numerically-optimized shaped pupils\cite{2022SPIE12180E..5KN}. As described below in more detail,  the project was organized in three levels of milestones, from open-loop natural ambient conditions,  to closed-loop control under both natural and artificial drift conditions.  HiCAT has now reached and exceeded the performance for all three levels of milestones.  
 
\section{Testbed overview and Summary of project goals}

\subsection{Project goals and milestones}

The last few years of HiCAT were supported mainly under the NASA Strategic Astrophysics program (Technology Demonstrations for Exoplanets Missions). This programs requires the definition of formal milestones, which were outlined in a white paper at the beginning of the project  \cite{TDEM-white-paper}\footnote{\url{https://exoplanets.nasa.gov/internal_resources/1186/}}.   

The objective was to advance the technology readiness levels (TRL) of segmented telescope and coronagraph systems for future terrestrial planet direct imaging missions, now known as HWO.  The stability of the wavefront delivered from the telescope presents special challenges to the coronagraph in order to  achieve a factor of $\sim10^{-10}$ starlight suppression. The telescope and coronagraph must be considered together and their technologies advanced as an integrated system.  The goal of HiCAT was to advance this these system-level aspects to TRL-4, i.e. in ambient laboratory conditions, before moving onto more advanced experiments with segmented-aperture coronagraphic systems in relevant vacuum environments, in order to reach higher TRLs (5-6). 

The three milestones to meet the TRL-4 requirements are based on the demonstration of  $10^{-7}$ contrast with an APLC in a 360 degree Dark Hole (DH) extending from $4.5 - 12\lambda /D$, in a 6\% band centered at 638 nm. The first milestone corresponded to static contrast under natural drift.  The second milestone demonstrates closed-loop high-contrast wavefront control in the presence of static segment misalignments, in particular by the operation of a LOWFS to sense and correct segment misalignments in the telescope simulator. The third milestone adds a dynamic component with artificial drifts added to the telescope simulator.  In consistency with other TDEMs \cite{SerabynWP10,SerabynWP14}, all contrasts reported here are defined as the simple average contrast in the entire annular DH (average intensity in the DH normalized by the peak intensity of the direct PSF), and reported at a confidence level of $90\%$.  Also, all reported contrast are the results of standalone scripted experiments starting from a state without any wavefront control on the deformable mirrors. 

 The milestones also included  comparisons demonstrating agreement between experimental data and analytical, or end-to-end, numerical modeling. The final results we present in these proceedings for our SAT milestones comfortably exceed the performance requirements with static, LOWFS natural drifts, and artificial drifts from our SAT milestones  \cite{TDEM-white-paper}\footnote{\url{https://exoplanets.nasa.gov/internal_resources/1186/}}.

\subsection{Testbed hardware and operations}

The HiCAT testbed includes a segmented aperture telescope simulator (37-segment IrisAO) where each segment is controllable in piston tip-tilt. Because the surface of IrisAO mirrors is reflective beyond the segments themselves, HiCAT defines its entrance pupil using an aperture stop that limits the beam to the contour of the outer segments (slightly undersized), therefore creating a non-circular, fully segmented and active telescope simulator.  This adds the ability to include demonstration of high-contrast in the presence of co-phasing wavefront errors,  and also to introduce temporal drifts for dynamical studies.

\begin{figure}[th!]
\includegraphics[width=1.0\textwidth]{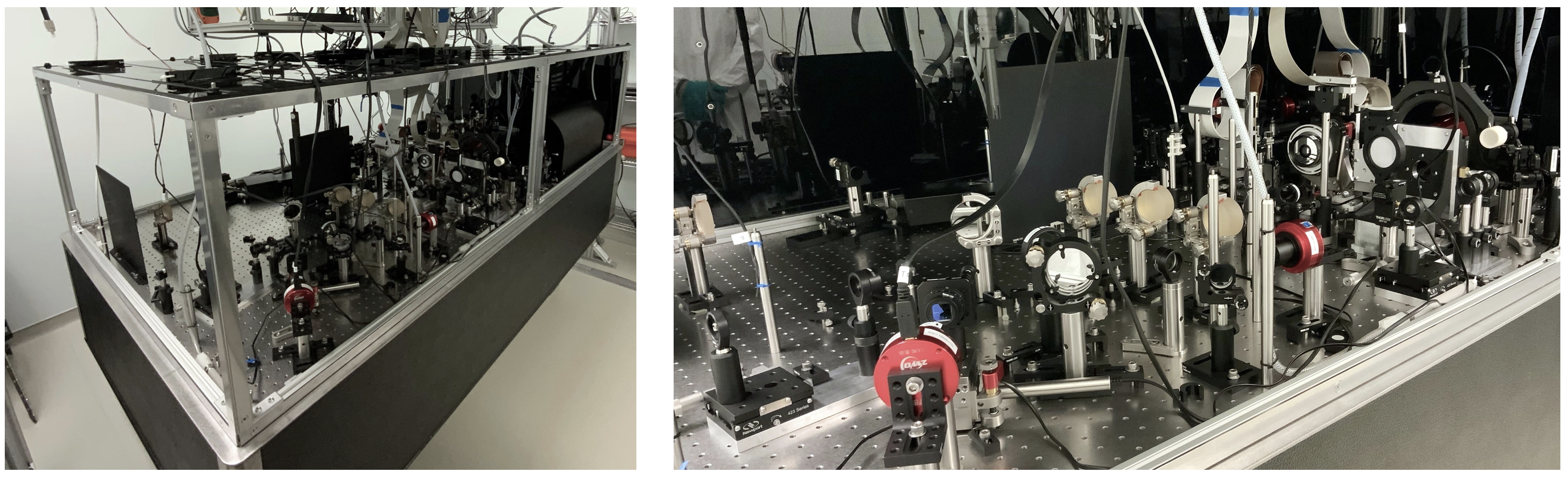}
\caption{\small Left: HiCAT testbed with its enclosure panels open. The testbed was entirely re-cabled and the top panels modified this year to reduce air gaps in the enclosure and improve stability. 
Right: detailed view of part of the testbed hardware, showing in particular the new Mid-Order Wavefront Sensor (MOWFS).  The Zernike masks can be seen (slight blue tint from its AR coating) next to the MOWFS camera at the bottom left.  The reflective apodizer can be seen in the middle of the testbed. }
\label{fig:picture_hicat_bench}
\end{figure}

The HiCAT testbed is shown in Figure \ref{fig:picture_hicat_bench}.  The instrument fits on an 8 x 4 ft optical table, and is covered by an enclosure for background light isolation and environment control. In 2023 the top panels were redone together with a full re-cabling of the testbed to reduce air gaps.  Since the DMs are sensitive to humidity, HiCAT operates under a constant dry air purge,  which has the effect of potentially increasing turbulence.  In order to minimize these effects as much as possible and also to mitigate humidity-related drifts (likely due to an hygroscopic adhesive), we added a closed-loop humidity control system, which maintains the testbed at 8\% relative humidity with $\sim0.1\%$ stability. These improvements have very significantly improved the stability of HiCAT compared to the condition of previously reported results. 

\begin{figure}[th!]
\center
\includegraphics[width=\textwidth]{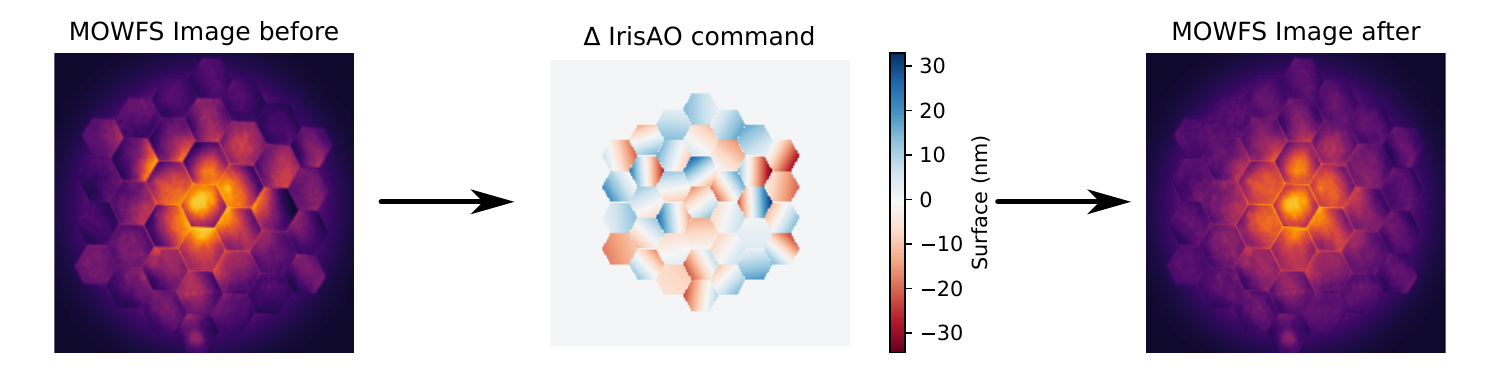}
\caption{\small  Previous results on HiCAT were obtained with open-loop calibration of the IrisAO based on Fizeau interferometric data. With the installation of the MOWFS we were able to calibrate independent piston/tip/tilt for the mirrors without the apodizer in the way, and without correcting these discontinuities with the continuous DMs. The MOWFS provides a full image of the entire 37-segment aperture and is therefore well suited for this calibration.}
\label{fig:iris_ao_calibration}
\end{figure}

Also this year, we implemented the Mid-Order Zernike sensor, which is visible in the right panel of Figure \ref{fig:picture_hicat_bench}.  The purpose of this MOWFS is to investigate multiple concurrent control loops as envisioned for HWO \cite{2022SPIE12180E..2KC}. The advantage of the MOWFS is that is has a full view of the entire segmented aperture before the application of the apodizer.  While  the segmented DM has very high open-loop repeatability it needs to be calibrated in its final location, and finally phased before performing high-contrast experiments (as would be the case on a real telescope system).  For routine operations we are using a dOTF calibration (differential Optical Transfer Function) \cite{Codona2013dOTF,2023SPIE12680E..2HN}.  However, because HiCAT also includes  two Boston Micromachines 952-actuator micro electro mechanical (MEMS) ``kilo-DMs", it is important to distinguish the alignment state of the segments from that of the continuous DMs.  The MOWFS offers the ability to calibrate the segment phasing directly, which helped identify non-negligible segment-level piston tip tilts that were present and corrected by the continuous DMs. This new level of fine alignment of the segments is shown in Figure \ref{fig:iris_ao_calibration}.

\begin{figure}[th!]
\center
\includegraphics[width=0.9\textwidth]{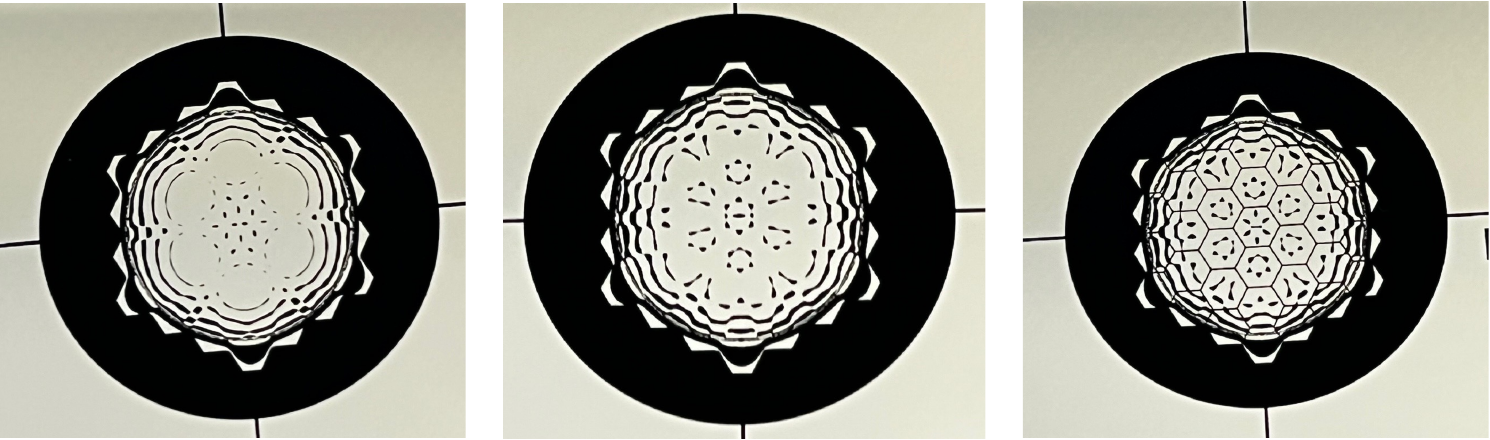}
\caption{\small  Pictures of the three reflective carbon nanotube apodizers corresponding to the latest optimization runs.  These apodizers were optimized specifically for the actual aperture by using a large format CMOS detector at the location of the apodizer itself. The apodizer used in this paper (left) is optimized for an off-axis non-circular aperture with 37 hexagonal segments and does not include segment gaps.}
\label{fig:CNT_apodizers}
\end{figure}

\begin{figure}[th!]
\center
\includegraphics[width=1.0\textwidth]{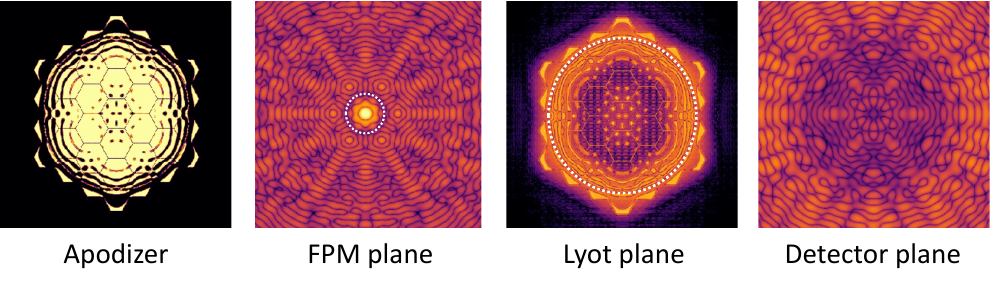}
\caption{\small Simulated illustration of the main coronagraphic planes.  From left to write: Apodizer image super-imposed with the segmented aperture (the apodizer does not include gaps). Focal plane image with FPM (white dotted circle).  Lyot plane showing diffracted light outside of the aperture, and Lyot Stop size for reference (white dotted circle).  Final coronagraphic image plane in the absence of wavefront error and without wavefront control.  The APLC design was optimized for $10^{-8}$ average dark zone contrast in 10\% broadband.}
\label{fig:aplc_each_plane_sim}
\end{figure}

Apodizers for HiCAT have been developed in partnership with \textit{Advanced NanoPhotonics}\footnote{\url{https://www.advancednanophotonics.com/}}\cite{2018SPIE10698E..1OS, 2022ApSS..57952250I,2010SPIE.7761E..0FH} using carbon nanotubes. 
 We previously reported an issue  \cite{2022SPIE12180E..26S}  with pupil alignment that currently prevents having an apodizer perfectly aligned with the segmented aperture. A re-alignment plan was studied but is no longer possible given limited funding, and we have instead implemented a mitigation strategy. Using a 24x36mm format CMOS camera, we have calibrated the actual aperture both at the conjugated plane and at the current location of the reflective apodizer, and have generated the optimal apodizers to reach $10^{-8}$ theoretical broadband contrast in 10\% band. Pictures of these carbon nanotube apodizers are shown in Figure \ref{fig:CNT_apodizers}. All results presented herein correspond to the first (leftmost) apodizer in a significantly defocused plane compared to the segmented aperture. This created challenges as it is extremely difficult to align the apodizer to the segmented aperture.  Our end-to-end optical model was used to confirm that this apodizer and defocus should not create noticeable issues at the current level of contrast on HiCAT and therefore we have concentrated efforts on this design.  These apodizers are combined with a circular Lyot Stop, and the main coronagraphic planes (without wavefront control) are shown in Figure \ref{fig:aplc_each_plane_sim}.  Also note that we are using an off-axis geometry, as the last cycle of design and manufacturing on this project followed the Decadal Survey recommendation for an off-axis configuration.  It has been very recently announced that NASA is considering again the possibility of an on-axis configuration as part of the GOMAP studies, and HiCAT could certainly be used in this mode again at a later time. 

The HiCAT software infrastructure was rebuilt in the last two years (E.H. Por, in preparation) to enable concurrent operations with multiple closed loops (DH electric-field conjugation, or stroke minimization at the same time as LOWFS control loops). Our hardware control library CATKit2 \cite{Catkit2} provides a service-oriented architecture with low-latency inter-process communication using shared memory. While the backend and communication layer are written in C++, the testbed can be operated and scripted fully from Python.   We use the Electric Field Conjugation (EFC) algorithm\cite{2007EFC} to create the DH with pair-wise estimation for electric field sensing with single-actuator probes (four pairs). 


\section{Milestone  1 results: TRL-4 component demonstration of APLC on a segmented aperture}

In previous SPIE proceedings, we have shown monochromatic results, using a laser diode at 638 nm.  Here using our tunable filter and supercontinuum laser, we generalize these results to narrowband of 20 nm at 680 nm (i.e. 3\%), and broadband (6\% and 9\%).
We have also moved the central wavelength to 680 nm from 638 nm as originally the case when using monochromatic laser diode light.  This small shift in central wavelength has no impact on the validity of the results, and was decided to optimize for performance including combined chromaticity of broadband source spectrum, pinhole lens chromaticity, Neutral Density filter transmission and detector response.  

\subsection{Narrowband APLC Results}

\begin{figure}[th!]
\includegraphics[width=1.0\textwidth]{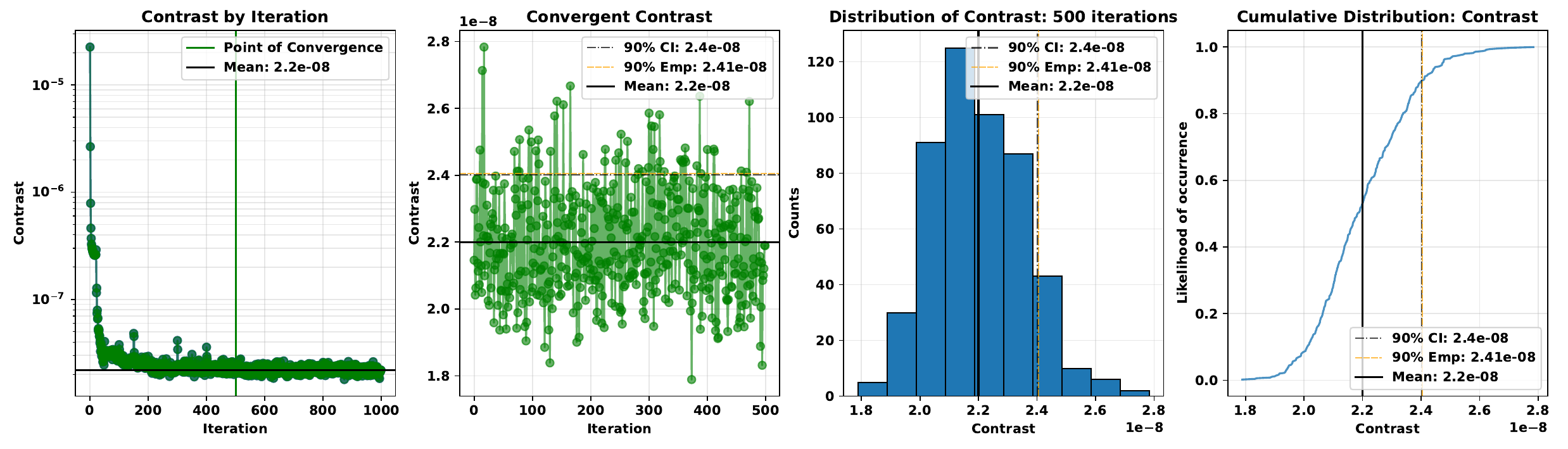}
\caption{\small  Standard monochromatic performance as defined by the SAT-TDEM HiCAT white paper.  First panel: Contrast convergence as a function of iteration number. The data set to the right of the vertical line at 500 iteration corresponds to the convergent contrast (second panel).  The third and fourth panel detail the convergent contrast statistics. These results correspond to standard performance on HiCAT using automated scripts.}
\label{fig:monochromatic_contrast}
\end{figure}

We show an example of a standard DH digging experiment in Figure \ref{fig:monochromatic_contrast}. Starting from the open-loop flat wavefront calibration of the Boston DMs, we first proceed to a fine target acquisition to center the PSF on the focal plane mask using the Target Acquisition (TA) camera.  We then proceed to a fine wavefront calibration using the dOTF algorithm, which then corrects any residual phase error (either coming from segments misalignments or from the general continuous alignment state of the testbed)  using the in-pupil continuous DM. As this calibration may introduce a very small amount of tip-tilt, we proceed to a second, fine-alignment TA.  The final step is the Lyot stop alignment.  This process guarantees fully automated, reproductible results. A standard experiment consists of 1000 iterations with at least 500 iterations past convergence.  The EFC loop continues to run during the convergent contrast section, and we define the 90\% confidence level contrast from the statistics of the convergent contrast.  Typical narrowband (3\%) contrast is $2.5\times 10^{-8}$.  

\subsection{Broadband APLC results}

\begin{figure}[th!]
\includegraphics[width=1.0\textwidth]{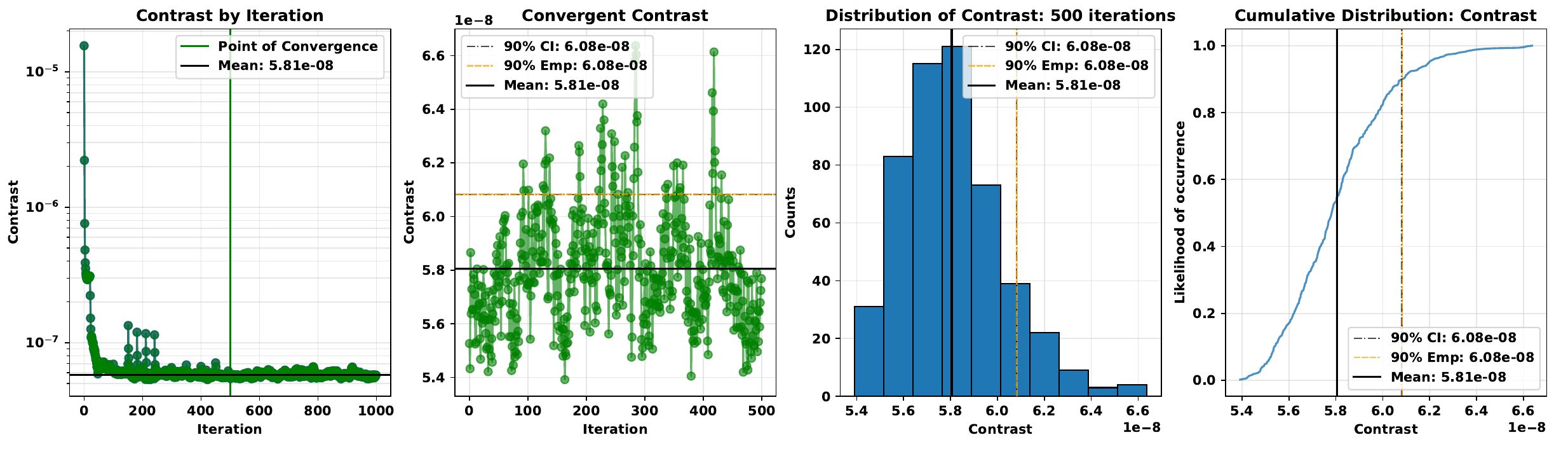}
\caption{\small  Standard $\sim9\%$ broadband performance similar to the narrowband results shown in Figure \ref{fig:monochromatic_contrast}. For broadband experiments, sensing is applied sequentially to three wavelengths using our tunable filter light source. Broadband operations are therefore slower by a factor 10 compared to single-band operations, therefore more susceptible to stability drifts on these timescales.}
\label{fig:bb_results}
\end{figure}

In this section we present broadband results centered at 680 nm with a bandpass of 60 nm, i.e. $\sim9\%$.
Our sensing and control strategy in broadband follows a standard process that has been implemented on other testbeds \cite{2019SPIE11117E..1VS,2020SPIE11443E..2QL}. Broadband dark holes are obtained using a NKT VARIA tunable source connected to a super continuum broadband laser source. We use three sensing wavelengths (660 nm, 680 nm, 700 nm) each with a 20 nm (3\%) bandpass, therefore totaling 60 nm bandpass.  Scanning each wavelength sequentially, we perform a pair-wise electric field sensing using four pairs of single actuator probes to estimate the electric field at each wavelength.  Using the model control Jacobians at each of these wavelengths, we reconstruct a correction that is applied to the Boston DMs.  The segmented DM remains static in open loop throughout the experiment.  Similar results to the narrowband case are presented in Figure \ref{fig:bb_results}, where the typical contrast is $\sim 6 \times 10^{-8}$.

\begin{figure}[th!]
\center
\includegraphics[width=0.65\textwidth]{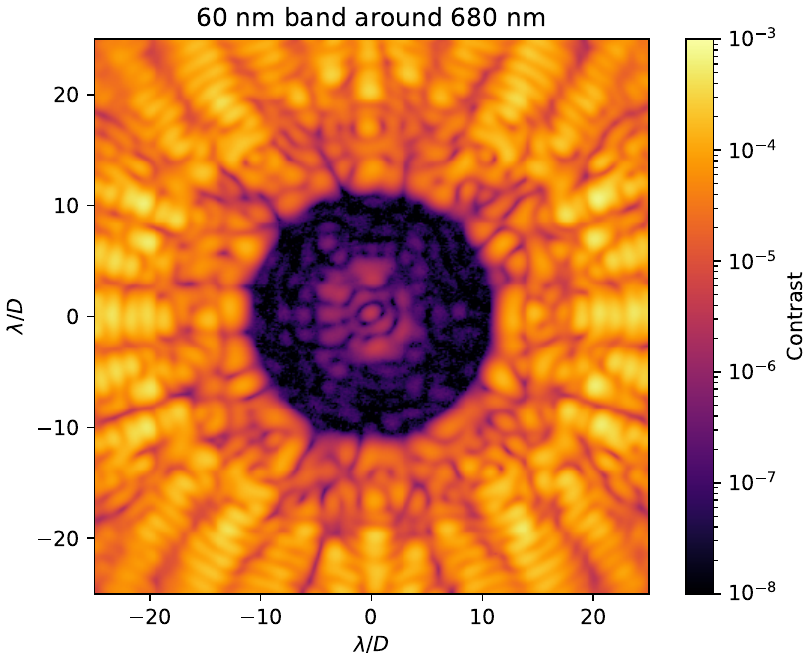}
\caption{\small  Truly broadband coronagraphic image obtained by adjusting the variable source filter to $\sim 9\%$ at the end of the standard broadband run shown in Figure \ref{fig:bb_results}.  The standard APLC dark hole in use currently has an IWA of $4.4 \lambda/D_{pupil}$ and OWA of $11 \lambda/D_{pupil}$. }
\label{fig:bb_dark_hole}
\end{figure}

In Figure \ref{fig:bb_dark_hole} we show the actual broadband dark hole by holding the last correction in open loop and switching the source to a full 60 nm bandpass, therefore capturing the actual broadband image in single exposures.  The spectrum of the source has a lot of oscillations, which is noticeable in the radially elongated speckles. 

\begin{figure}[th!]
\includegraphics[width=1.0\textwidth]{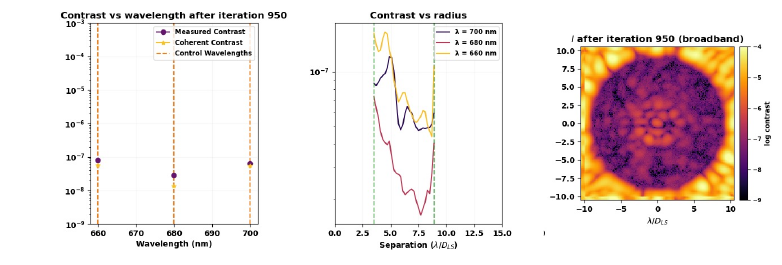}
\caption{\small  Average contrast in the DH for each of the three 20 nm bands used to control the full 60nm bandpass corresponding to the results shown in Figure \ref{fig:bb_results}. \textit{Center:}  Radial cuts of contrasts (azimuthal average) for each wavelength, showing significant discrepancy at short separation.  This is due to model mismatch in broadband, which is currently limiting performance. }
\label{fig:bb_wavelength_performance}
\end{figure}

In Figure \ref{fig:bb_wavelength_performance} we show the average DH contrast at each wavelength (left), and the corresponding radial profiles (azimuthal averages). While the integrated contrast is relatively flat across the full 60 nm band, the radial contrast plots exhibit significant mismatch at short separation.  We have attributed this issue to model mismatch, which is currently limiting the broadband APLC (contrast performance in 9\% is at least two times worse than in 3\% band). 

\subsection{Results with the Phase-Apodized Pupil Lyot Coronagraph (PAPLC)}

HiCAT has the ability to support a several variants of Lyot coronagraphs designs, and we also show here results for the PAPLC  \cite{Por2020PAPLC,por2021SPIE,10.1117/12.2677757}. This design uses a knife edge focal plane mask, which allows for very small IWA at the expense of a half dark hole.  PAPLC does not include any amplitude apodizer and therefore is a simpler implementation on HiCAT.  A phase apodization is used instead on the continuous DMs.  We show broadband contrast in  $9\%$ band of $4.2\times 10^{-8}$ ($2-13\lambda/D_{pup}$, half DH) with PAPLC and narrowband contrast of $2 \times 10^{-8}$, limited mainly by ambient conditions at short separations \cite{2022SPIE12180E..26S} (see also Figure \ref{fig:psd_tip_tilt}).  Large broadband DH (25\%) can also be obtained albeit at reduced contrast performance. 

\begin{figure}[th!]
\includegraphics[width=1.0\textwidth]{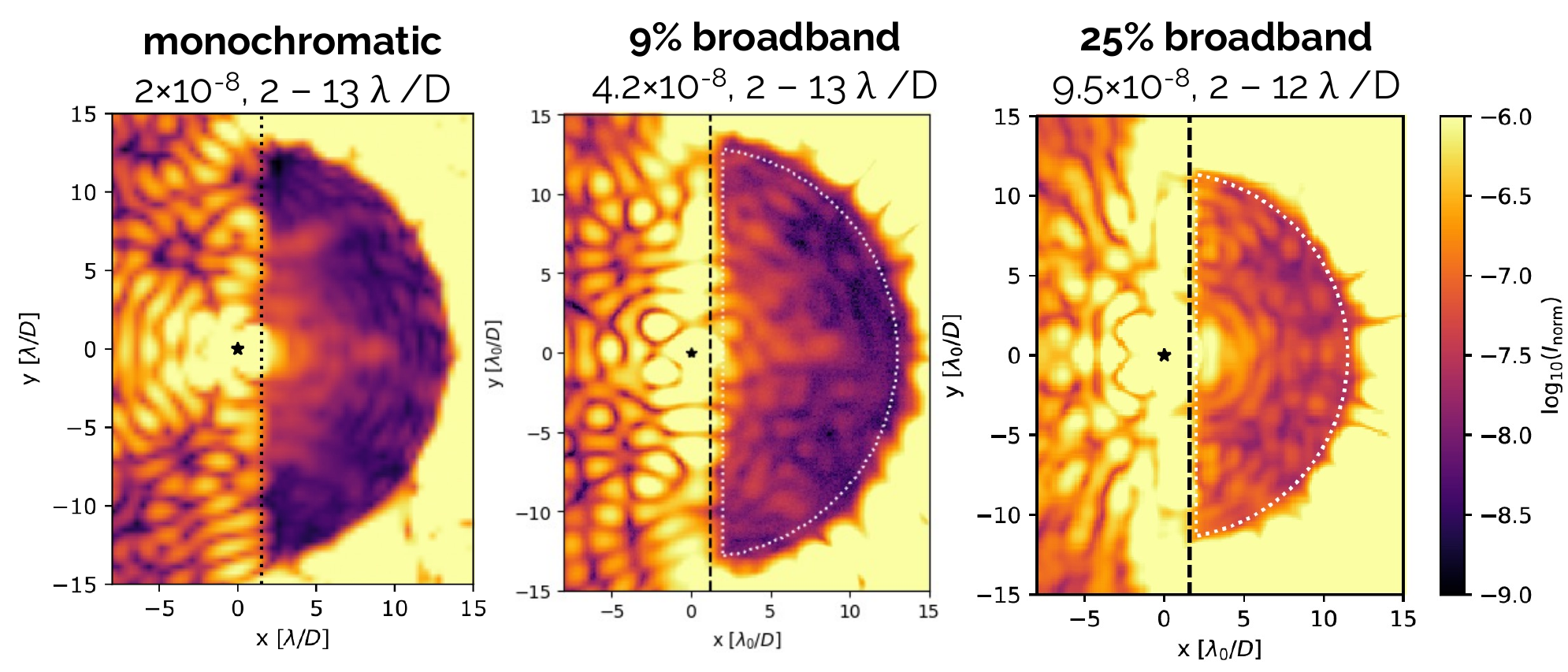}
\caption{\small  PAPLC monochromatic and broadband results, showing half DH with very aggressive IWA ($2\,\lambda/D_{pup})$. In the narrowband case and with the small IWA the performance is naturally limited by environment conditions. These results implemented a correction of tip-tilt for the 80 Hz peaks using the continuous DMs (see also Fig.\ref{fig:psd_tip_tilt}).}
\label{fig:bb}
\end{figure} 

\section{Milestone level 2a/2b: Dark hole with CLOSED-LOOP CONTROL UNDER NATURAL and Artificial DRIFTS.}

\subsection{Characterization of testbed natural drifts and turbulence}
\begin{figure}[th!]
\center
\includegraphics[width=0.85\textwidth]{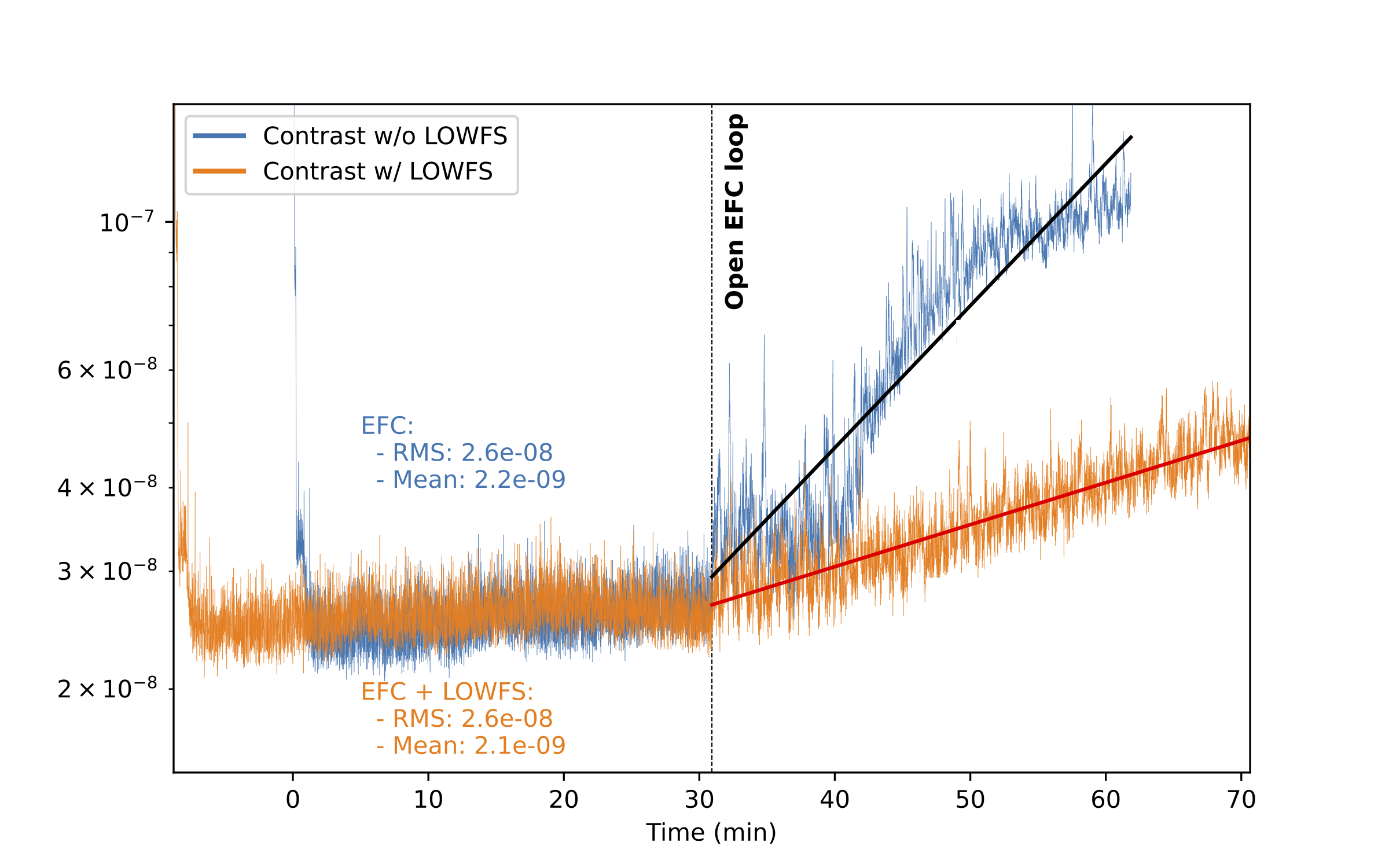}
\caption{\small  Characterization of open loop drifts in narrowband light (3\%) in a $\sim 1$ hour experiment. The first half corresponds to EFC closed loop (with and without LOWFS) and the second half (about 30 min) to open-loop EFC drifts with and without LOWFS stabilization.  The doubling time is respectively 48 and 14 minutes, demonstrating the utility of the LOWFS on the stabilization of long-term low-order drifts.  The closed-loop dark hole experiments presented here are typically of the order of minutes and therefore not significantly impacted by long timescale drifts. }
\label{fig:long_term_stablity}
\end{figure}

 Before investigating the contrast performance of the APLC with LOWFS under both natural and artificial drifts, we investigate the testbed stability.  Figure \ref{fig:long_term_stablity} shows the contrast evolution over about an hour both with and without LOWFS stabilization.  At the half-way mark (30 min) the EFC control loop is open and the testbed is drifting naturally.  We observe a doubling time respectively of 48 min and 14 min with and without LOWFS stabilization.  This confirms the expectation that long-term drifts have a significant low-order contribution, which is consistent with our observation of temperature and humidity dependance of the testbed.
Given the times scale of our dark zone experiments (closed-loop EFC at 5 Hz single band and 0.5 Hz with three wavelengths), the open-loop slow drift rate is not a concern at the current HiCAT levels of contrast ($\sim 10^{-8}$ and above), especially in closed loop. 

\begin{figure}[th!]
\center
\includegraphics[width=0.75\textwidth]{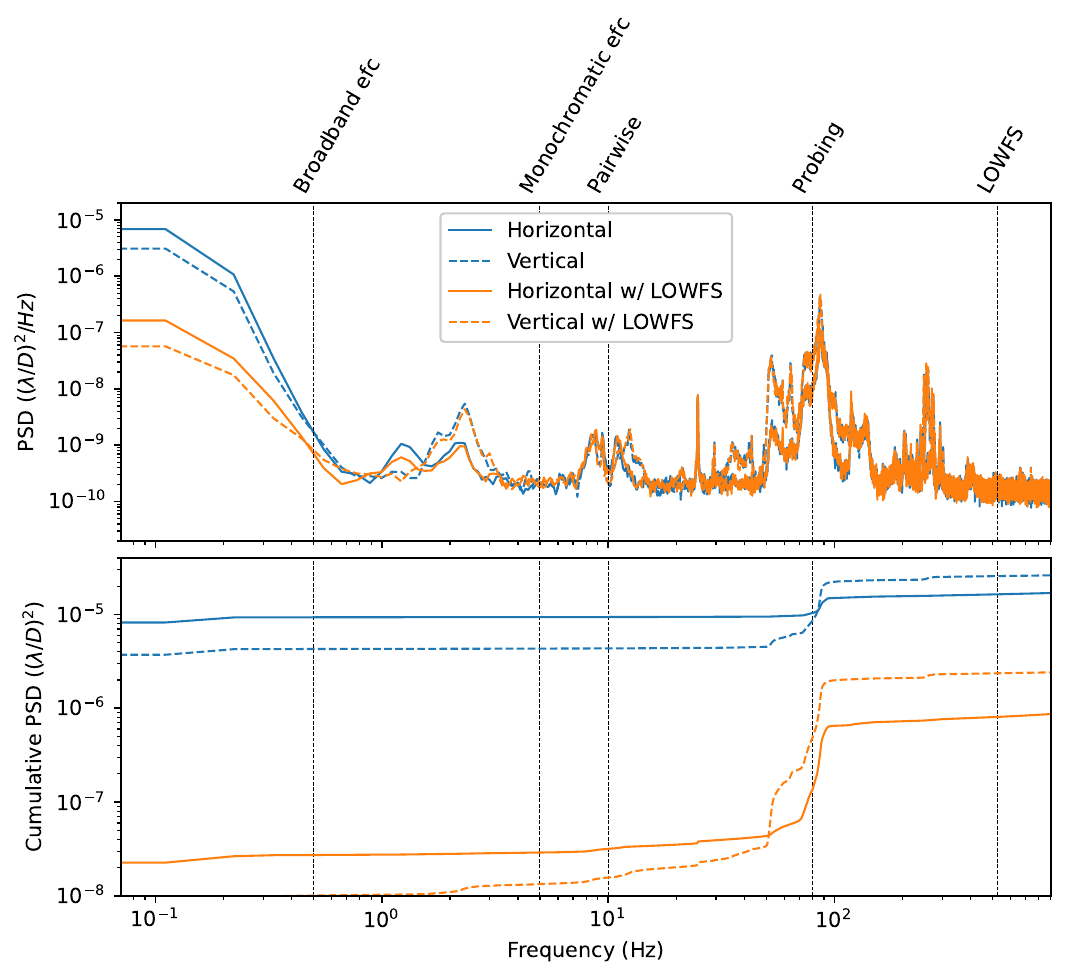}
\caption{\small  Top: temporal PSDs of the tip-tilt acquired on the target acquisition camera at 1800 Hz over 500s.  The frequencies of all relevant sensing and control loops are shown with vertical lines. Bottom: Cumulative PSD  corresponding to the variance of the tip-tilt. Note that the broad peaks around 80 Hz contribute to the most energy (due to the log-log scale). The LOWFS has very noticeable effect at low frequency, however the LOWFS control has not yet been optimized and is a simple integrator with very low gain (0.01). Broad peaks around 80-100Hz due to acoustic vibration are at the probing frequency and therefore they impact sensing and are not corrected in closed loop. The 10x slower speed in broadband due to filter wheel moves might therefore also contribute to the lower performance. }
\label{fig:psd_tip_tilt}
\end{figure}

We then investigate the effect of faster disturbances using tip-tilt measurements as a proxy. We acquire 1800 Hz direct images to measure their tip-tilt over $\sim 10$ minutes. The PSD is shown in Figure \ref{fig:psd_tip_tilt}.  Turbulence and vibrations are impacting the sensing loops. While this is impacting all our results, this is only noticeable in narrowband at short separations (in particular PAPLC narrowband results included 80 Hz tip-tilt correction for best performance). The LOWFS control loop has not been optimized yet (as evidenced by the noise amplification above a few Hz), and this will be a subject of future work.  The 80 Hz energy has been associated with acoustic vibration from the cleanroom fans, and could be potentially be mitigated further.  Unfortunately while turning off the fans removes these components, the lab environment drifts rapidly in temperature and humidity, which is counter productive for achieving best results. 

\subsection{Results with artificial low-order drifts and static segmented aperture.}

\begin{figure}[th!]
\includegraphics[width=1.0\textwidth]{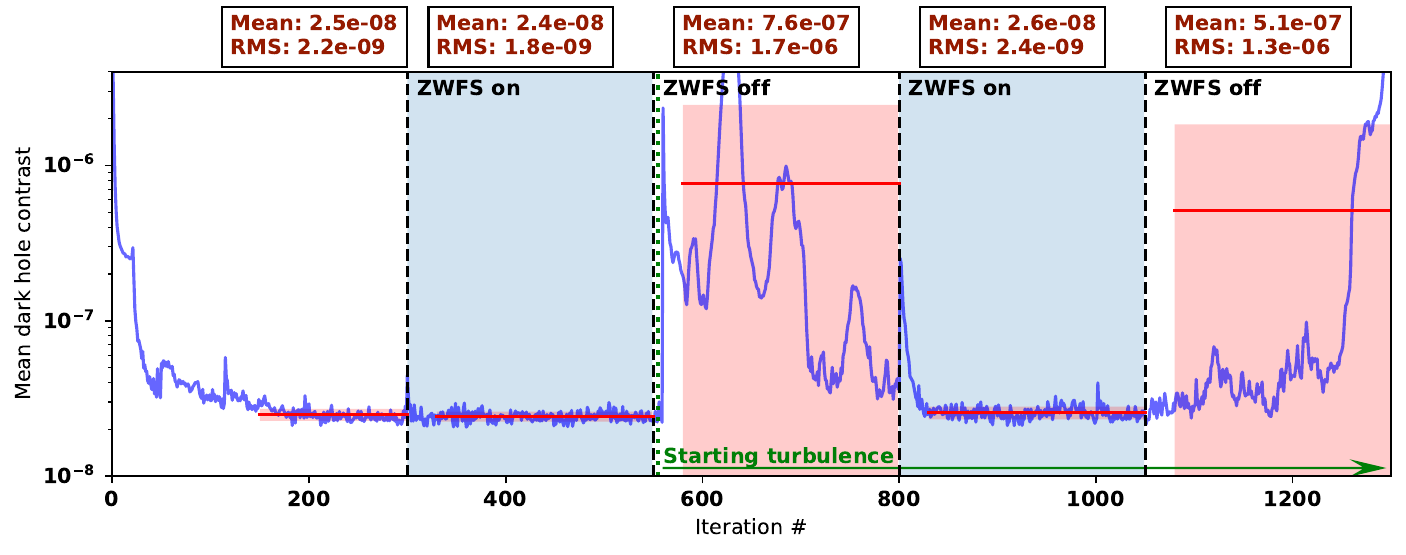}
\caption{\small  Monochomatic experiment with LOWFS and under natural and artificial low-order wavefront drifts.  The first section (iteration 0-250) corresponds to the initial DH digging from flat DMs, under natural conditions and without LOWFS.  The LOWFS is then turned on under natural drift conditions (250-500), showing slightly improved contrast mean and RMS. Artificial low-order drifts of  12 nm RMS, with 6 injected Zernike modes and 10 modes corrected (calibrated on Zernike modes) are then added (green arrow).  The experiment corresponds to a total duration of about 100 seconds and total decorrelation is achieved in 30s.  The LOWFS gain is  0.01. With ZWFS under drift the contrast baseline is fully recovered, albeit with higher variance. }
\label{fig:low_order_drift_narrow}
\end{figure}

We first investigate the case of a static segmented DM and injection of low-order drifts on the continuous DM, in order to validate the LOWFS.  The results are shown in Figure \ref{fig:low_order_drift_narrow} in narrowband.  In the second panel of this figure, we obtain the best contrast ($2.4\times 10^{-8}$) with concurrent EFC and LOWFS closed loops under ambient drifts.  The RMS of $1.8\times 10^{-9}$ is slightly reduced with closed loop LOWFS and this is consistent for a large number of runs. Even in the presence of ``large" low-order random drifts (12 nm RMS surface error), closing the loop recovers the contrast performance with a slightly increased RMS. 

\subsection{Results with segmented aperture drifts}

In this section we investigate the case of random walk drifts in piston, tip and tilt on the segmented mirror, in addition to low-order drifts as in the previous section. 

\subsubsection{Limitations of segmented telescope simulator resolution for drift studies}
\begin{figure}[th!]
\center
\includegraphics[width=0.8\textwidth]{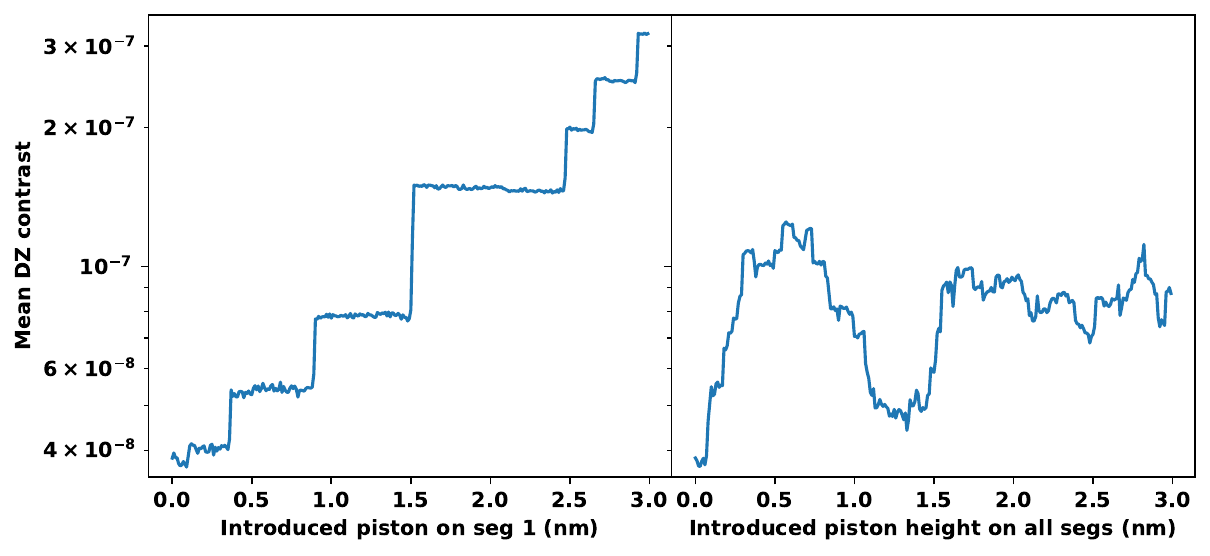}
\caption{\small  Left: single segment piston showing the multiple steps as each of the three legs trigger before forming a full piston.  Right: pushing a global piston on the DM, first degrading the contrast as segments leg first trigger, and then recovering the contrast when the full mirrors has moved by a full piston step. While limited in resolution, our segmented mirror is adequate for studying dynamic perturbations at the current HiCAT contrast levels.}
\label{fig:iris_quantization}
\end{figure}

Our telescope simulator relies on the IrisAO 37 segment PTT111L, and we investigate the impact of our 14 bit electronics quantization steps on the contrast. This effect was studied in more detailed in another contribution\cite{Buralli2024}. In Figure \ref{fig:iris_quantization} we illustrate both the impact of a single segment piston ramp and a global piston ramp.  Because each segment has 3 legs with independent quantization steps, a requested ramp in piston amplitude on a single segment will have the first two legs trigger tip/tilt, and finally the third leg to get to a full piston step. When applying a ramp of global piston on all segments, the contrast degrades significantly as segment legs move steps, before improving again to a local minimum corresponding to a flat aperture (all segments moved by one piston step). 
While this effect is limiting our ability for studying continuous drifts, this proves adequate resolution for studying segment disturbances corresponding to open loop DH contrast in the $10^{-7}$ to $10^{-6}$ range. 

\subsubsection{Stabilization of combined low-order and segment drifts using EFC and LOWFS in 6\% broadband}

Because of our sequential operations for broadband DH control (sensing the E-field in each wavelength and calculating the broadband correction), it is challenging to implement the LOWFS in broadband\cite{10.1117/12.2677790}. Indeed, since the beam of light is interrupted during each filter wheel move, the LOWFS loop must be opened and closed each time. Moreover, each wavelength has a different reference in the LOWFS, therefore requiring recovery time for the loop to catch up at each wavelength switch.  While we have successfully implemented this strategy, it is difficult to stabilize fully.  More importantly, this control strategy is irrelevant for an actual mission, since sequential sensing will be replaced in flight by an Integral Field Spectrograph or an Energy Resolving detector \cite{Steiger2024}.  We therefore explored a single band correction in a reduced bandpass of $6\%$ (our minimum SAT milestone requirement).  Of course this does not provide a true broadband correction,  but an average monochromatic correction over the full band. This ``broadband uniband" control strategy still provides good results in $6\%$ comparable to the 9\% sequential control, but not in $9\%$ uniband where the contrast is above $10^{-7}$. 

\begin{figure}[th!]
\includegraphics[width=1.0\textwidth]{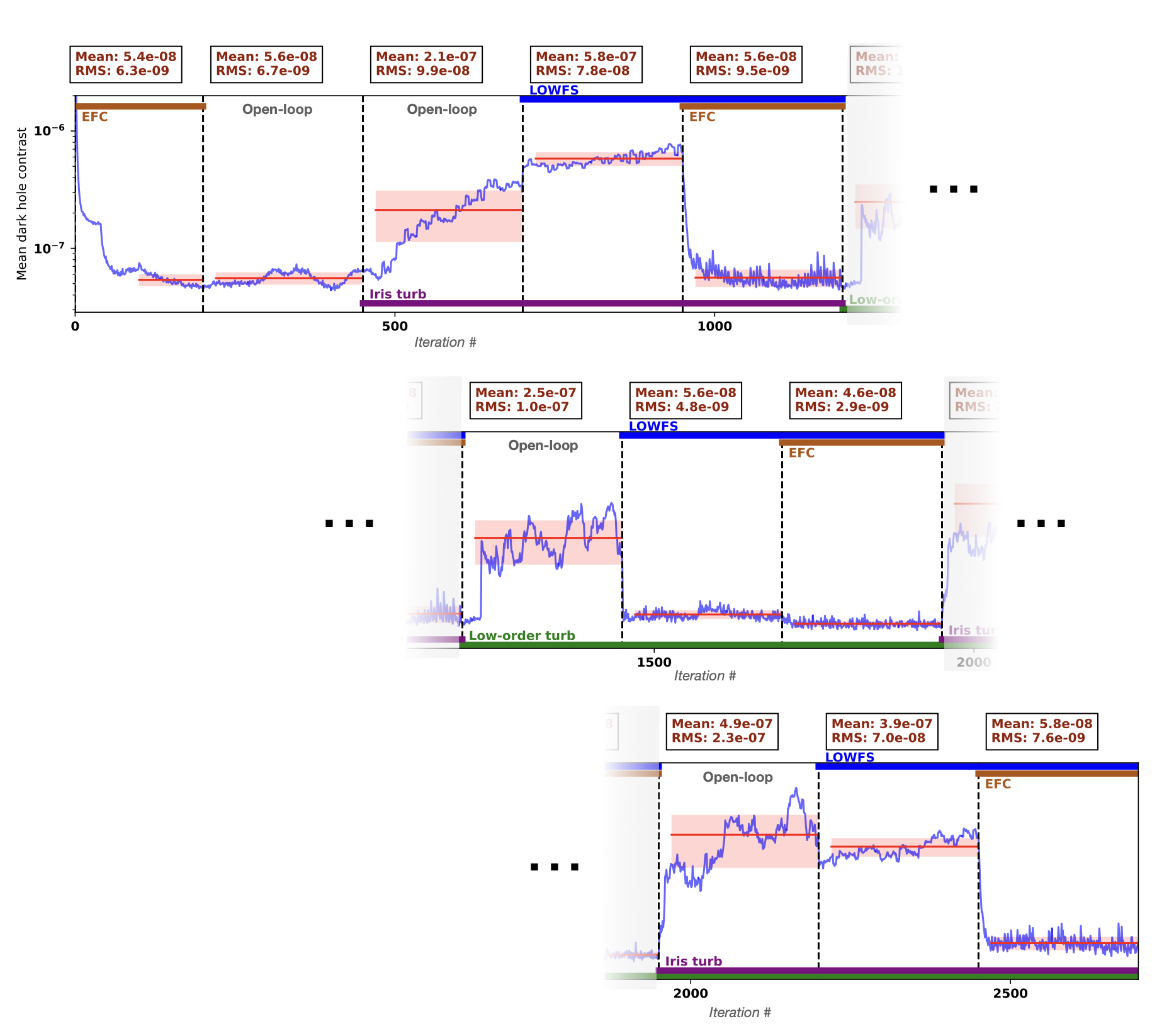}
\caption{\small  Dynamic experiments with artificial drifts in 6\% broadband with both segment and low-order drifts, and with concurrent closed loops (EFC and LOWFS).  Being in as single band, the experiment runs at the same speed as narrowband (i.e. 5 Hz).  First row: initial DH digging and full open-loop contrast showing indication of underlying stability in this mode, followed by segment-level drifts with a random walk of 100 pm / $\sqrt{s}$ (piston) and 50 nrad / $\sqrt{s}$  (tip/tilt).  Second row: low-order turbulence only with constant 7 nm RMS surface error spread over 6 Zernike modes and with a total decorrelation of 30 s (i.e. corresponding to 150 iterations).  Third row: simultaneous low-order and segment drifts.   In each of these case, we show three panels, first open loop (only the drift component) followed by LOWFS alone and finally concurrent LOWFS and EFC. The LOWFS is as simple integrator with a gain of 0.01 and runs at 500 Hz, correction is applied on 6 modes. }
\label{fig:iris_drifts}
\end{figure}

This experiment is illustrated in Figure \ref{fig:iris_drifts}, where we successively apply segmented and low-order drifts first alone, then combined.  We also show in each case successively open-loop, LOWFS, and concurrent LOWFS and  EFC. Our service-oriented architecture enables asynchronous loops \cite{Catkit2,2022SPIE12180E..26S} with independent channels on the deformable mirrors. 

 In the case of segment-level drifts, the LOWFS alone does not recover contrast but reduces the general slope by stabilizing the global low-order modes on the mirror. EFC can then recover the contrast level, albeit in this case with significantly increased RMS. The behavior is different with low-order modes only as the LOWFS can in this case recover the contrast level;  EFC provides slight further improvement. With both low-order and segment drifts, the LOWFS recovers part of the contrast first, then EFC. 

When running a 9\% uniband correction, the contrast is limited above $10^{-7}$. However,  the Zernike LOWFS works well in that 9\% band and can stabilize the contrast with about 30\% increase in RMS, on par with the bandpass. Unfortunately we cannot test the LOWFS at 9\% at our best broadband contrast.  

\clearpage
\section{Numerical and Analytical modeling}

\begin{figure}[th!]
\center
\includegraphics[width=0.72\textwidth]{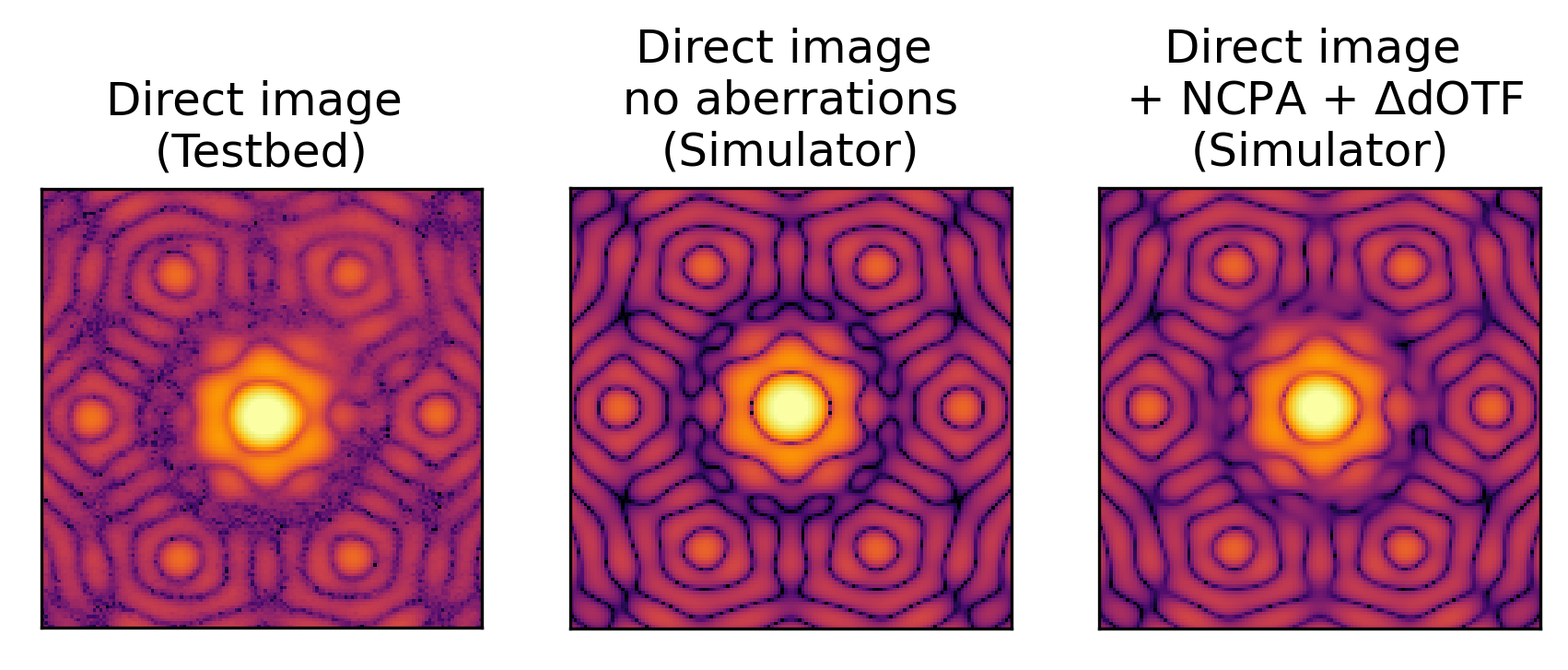}
\caption{\small  Calibration of upstream (before FPM) and downstream (after FPM) wavefront aberrations. HiCAT has a dedicated phase retrieval camera which offers a proxy location for the FPM, via a high-quality fold mirror. Using dOTF phase retrieval at both this location and at the science camera, we can calculate the non-common path aberrations that corresponds to the downstream wavefront errors.  This illustration compares the data with the direct image without calibration (note in particular the absence of trefoil, which is properly calibrated by measuring residual aberrations at the phase retrieval camera and adding the downstream aberrations).}
\label{fig:model_ncpa}
\end{figure}

\begin{figure}[th!]
\includegraphics[width=1.0\textwidth]{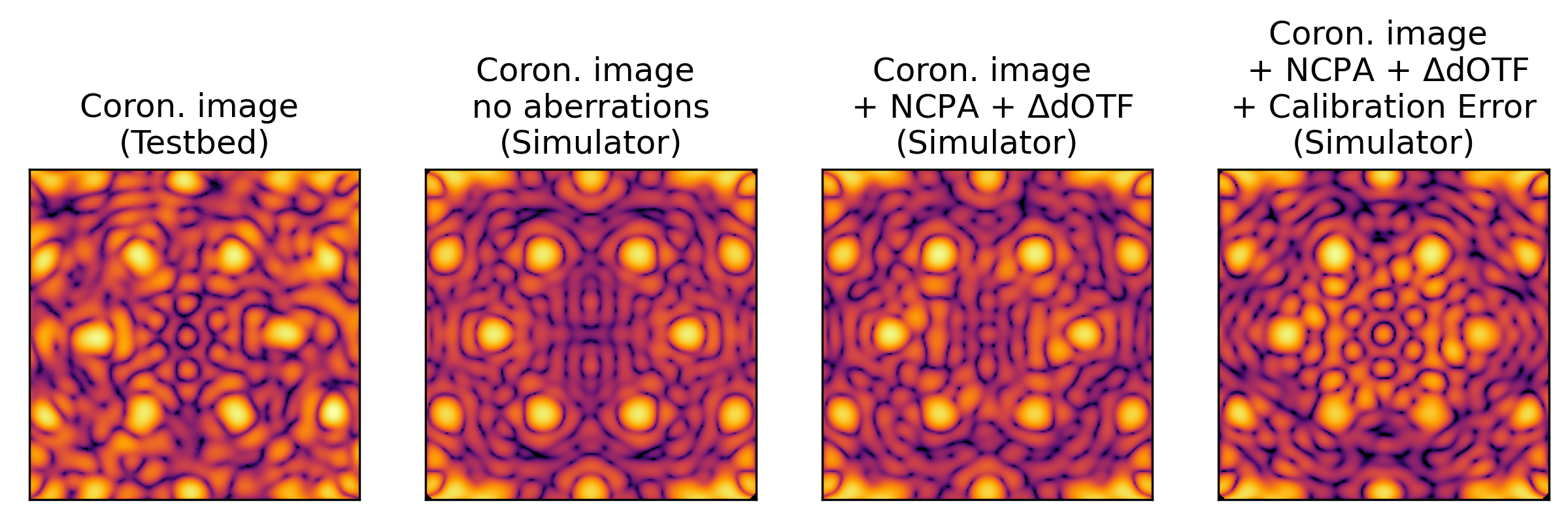}
\caption{\small  Modeling experimental data from the testbed (left). We have implemented an optical model, which we use to calculate the control Jacobian.  The first part of the model calibration effort is to match all the geometric parameters, (as-built magnifications, alignments etc.) in the absence of modelled errors (second panel).  This model is what we are using to calculate the control Jacobian. Matching fine structures in the image is much more challenging as multiple effects are degenerate. The third panel shows a more realistic data-based calibration including NCPA between upstream and downstream aberrations, and with residual estimation of hardware wavefront error ($\Delta dOTF$).  The inner 6 spots around the center spot are currently challenging to understand and for example can be enhanced by adding defocus on the apodizer or LS beyond what is currently measured on hardware.  High-fidelity model matching of coronagraphic images in a complex system like HiCAT remains a challenging task. }
\label{fig:model_coron}
\end{figure}

The HiCAT numerical simulator implements all essential optical components (e.g., masks, DMs, cameras) but without including every single optical surface of the testbed (e.g., off-axis parabolas and lenses). This numerical model has been calibrated with hardware data to reproduce all detectors scales, and is used for generating the Jacobian control matrices.  We keep developing the model by adding more high-fidelity features to solve remaining model mismatch with the testbed data.  We give two examples of modeling results in Figure \ref{fig:model_ncpa} and \ref{fig:model_coron} respectively for direct and coronagraphic images.  For direct images, we have the ability to include aberrations downstream and upstream of the coronagraph mask, and we calibrate these non-common path errors by performing dOTF phase retrieval at both our phase retrieval camera (a proxy location of the FPM) and at the science camera \cite{2023SPIE12680E..2HN}. 
Accurate modeling of coronagraphic images is more challenging as these images are more sensitive to small changes, and many causes have degenerate effects.  We give an example of successive increased complexity in our model with residual phase errors, non-common path and apodizer defocus to show evolution of the structures.  Work will continue to keep developing this simulator into a high-fidelity model. 

\section{conclusions}

Over the past 10 years the HiCAT project has been contributing to maturing system-level technology for the future Habitable Worlds Observatory.  It is still, to this date, the only system-like testbed operational with a segmented aperture telescope simulator. 
In these proceedings we presented the final results obtained under our NASA Strategic Astrophysics Technology funding which was articulated around three milestones to demonstrate TRL-4:  static contrast in ambient conditions, and dynamic contrast under LOWFS stabilization under both ambient and artificial drifts. All performance metrics stated in the original milestones have been met and exceeded. 

At the moment, the narrowband contrast performance is limited by a combination of hardware and environment effects (DM quantization noise, detector effects, turbulence, and vibrations).  Our broadband contrast is still at least 2 times worse than our narrowband results which is due to model mismatch mainly, as the other effect are less significant.  Our simulation and modeling infrastructure is complete, but still a work in progress as we keep investigating the root causes of our model mismatch in broadband. 

Another indication of environment limitation is that contrast results are typically deeper when running LOWFS and EFC in narrowband. Since the LOWFS is not designed for contrast improvement and just stability, this means that the LOWFS helps the EFC loop by stabilizing drifts on the relevant sensing and control timescales. 

All the results presented here were in the off-axis telescope configuration, based on the recommendation of the 2020 Decadal Survey, but it would be desirable to continue investigations on HiCAT and especially in the area of on-axis architectures as they have become under re-consideration for HWO.  

\acknowledgments 

The HiCAT testbed has been developed over the past 10 years and benefitted from the work of an extended collaboration of over 50 people. This work was supported in part by the National Aeronautics and Space Administration under Grant 80NSSC19K0120 issued through the Strategic Astrophysics Technology/Technology Demonstration for Exo-planet Missions Program (SAT-TDEM; PI: R. Soummer), and under Grant 80NSSC22K0372 issued through the Astrophysics Research and Analysis Program (APRA; PI: L. Pueyo).
E.H.P. was supported in part by the NASA Hubble Fellowship grant HST-HF2-51467.001-A awarded by the Space Telescope Science Institute, which is operated by the Association of Universities for Research in Astronomy, Incorporated, under NASA contract NAS5-26555.
Sarah Steiger acknowledges support by STScI Postdoctoral Fellowship and Iva Laginja acknowledges partial support from a postdoctoral fellowship issued by the Centre National d’Etudes Spatiales (CNES) in France.


\end{document}